\documentclass{jpsj3}
\usepackage{txfonts}
\usepackage{graphicx}% Include figure files

\usepackage{braket}
\usepackage{here}
\usepackage{siunitx}

\title{Theory of simultaneous observation of photoelectron \\and emitted X-ray photon in X-ray emission spectroscopy \\and their entanglement}

\author{Ryo B. Tanaka\thanks{su23179r@st.omu.ac.jp} and Takayuki Uozumi}
\inst{Department of Physics and Electronics, Graduate School of Engineering, \\Osaka Metropolitan University,\\ Sakai, Osaka 599-8531, Japan} %\\

\abst{We have theoretically investigated the character of entanglement between photoelectrons and emitted X-ray photons in XEPECS process which is a coincidence spectroscopy of photoelectrons and emitted X-ray photons. In the calculations, we used a three-electron $sp$-model consisting of an inner-shell 1$s$ orbital and an outer-shell 2$p$ orbital. We found that entanglement occurs between spin of photoelectrons and linear polarization of emitted X-ray photons, using the wavefunction of the whole system including the photon as well as the electron system. Furthermore, we calculated the degree of entanglement and found that it shows maximum at a particular angle of emitted X-ray photon.}

\begin{document}
\maketitle

\section{Introduction}

Quantum entanglement is not only applied as a major resource in quantum information processing, but is also investigated to explore fundamental physics \cite{RevModPhys.81.865,Advancesentanglement,Entanglementcertification,PhysRevLett.119.240402}. Although there are numerous theoretical and experimental studies to investigate various aspects of quantum entanglement \cite{PhysRevLett.121.023602,BNCver2,energytimeentangled,Remotequantumentanglement}, most of these studies have been performed on entangled photons. Recently, electron-photon entanglement has gained importance due to its potential applications in long-range quantum teleportation and quantum repeater technology. For example, entanglement between electron spin and photon frequency/polarization were demonstrated in several experimental reports \cite{microphoton,electronphoton1,RevModPhys.94.041003}.

However, as pointed out by B.N.Chowdhury $et\ al.$ \cite{mainref}, in contrast to the low-energy regime, there are few reports on the entanglement behavior of high-energy particles. Recently, owing to the development in synchrotron radiation optics \cite{xrayoptics1,xrayoptics2,PhysRevLett.126.104802,xrayoptics4}, quantum optical experiments using high-energy X-rays, especially the generation of entanglement, have begun to become popular \cite{PhysRevLett.109.013602,PhysRevLett.119.253902,PhysRevA.97.063804,xrayregion4,PhysRevX.9.031033,10.1063/1.5092945,PhysRevResearch.1.033133,PhysRevLett.131.073601}. D. Jannis $et\  al$. \cite{10.1063/1.5092945} and S. Tanaka $et\ al$. \cite{doi:10.1143/JPSJ.61.4212} studied experimentally and theoretically the production of pairs of photoelectron and X-ray photon in X-ray emission spectroscopy processes.
However, while there is a study of spin-polarization entanglement in high-energy Compton scattering \cite{mainref}, quantum entanglement between photoelectron spin and emitted X-ray photon polarization due to X-ray inner-shell excitation has not been investigated.

In this context, the present work investigates theoretically the spin-polarization entanglement of photoelectron and X-ray photon in XEPECS, i.e., XES (X-ray emission spectroscopy) $\&$ cXPS (core-level X-ray photoemission spectroscopy) coincidence spectroscopy \cite{doi:10.1143/JPSJ.61.4212}. XEPECS is a spectroscopic method that simultaneously measures the kinetic energy of photoelectrons emitted by irradiations of incident X-ray photons and the energy of X-ray photons emitted through the radiative decay of inner-shell holes. 
In this paper, we treat the spin-, polarization-, and the X-ray emission angle- resolved XEPECS. Such a spectral process enables us to quantify the degree of spin-polarization entanglement. Moreover, we can investigate the emission angle dependence of it.
Needless to say, we focus on the process of photoemission and X-ray emission occurring at the same site in this study.
In a recent experiment, photoelectrons and X-ray photons from the same atom were observed by electronic excitation \cite{10.1063/1.5092945}. However, at present, XEPECS experiments have not been conducted probably due to technical difficulties, and thus, this study is positioned as a theoretical prediction. 

In this work, to investigate the fundamental aspect of entanglement, we use a simple atomic model consisting of two levels, inner-shell $s$ and outer-shell $p$ orbitals ($sp$-model). We quantitatively evaluate the degree of spin-polarization entanglement between photoelectrons and X-ray photons by XEPECS and show that it depends on the emission angle.

\section{Theoritical method}

In this study, we adopt a simple $sp$-model as an electron system in order to concentrate on the photoelectron-photon entanglement produced in the XEPECS process. The $sp$-model consists of inner-shell $s$ and outer-shell $p$ orbitals and has $s^{2}p^{1}$ electronic configuration in the initial state. The model may appear to correspond to B atom with $(1s)^{2}(2s)^{2}(2p)^{1}$ configuration, for example. However, we do not have the B atom in mind but adopt it as a minimal theoretical model to describe essential aspects of the photoelectron-photon entanglement produced in the XEPECS process. Note that, we denote the inner-shell orbitals as $1s$ and the outer-shell orbitals as $2p$, for simplicity. For a discussion on more realistic systems, such as $3d$ transition metal compounds, we need a more detailed calculation using, e.g., a charge-transfer cluster model of $\rm{MO}_{6}$-type \cite{doi:10.1143/JPSJ.69.1558} including full-multiplet effects in the transition metal atom. However, such a study will be the next step for us, and the purpose of this study is to reveal essential aspects of photoelectron-photon entanglement using a simple model.

The XEPECS process consists of photoelectron emission from the $sp$-model by irradiation of incident X-ray photons and the emission of X-ray photons due to the relaxation of the core-excited intermediate state. In order to describe the participating particles in the process, we consider the Hamiltonian,
\begin{align}
\label{1}
    H_{0}=H_{sp}+H_{PE}.
\end{align}
where $H_{sp}$ is for the $sp$-electron system and $H_{PE}$ the photoelectrons. In this study, we set the $sp$-electron Hamiltonian as
\begin{align}
\label{2}
    H_{sp}=\epsilon_{s}\sum_{\sigma}s_{\sigma}^{\dag}s_{\sigma}+\epsilon_{p}\sum_{m,\sigma}p_{m\sigma}^{\dag}p_{m\sigma}+H_{S-O}+H_{exch}.
\end{align}
In eq.(\ref{2}), $\epsilon_{s}$ is the energy of the inner-shell $1s$ state and $s_{\sigma}^{\dag}(s_{\sigma})$ is the creation (annihilation) operator for the $1s$ state with spin $\sigma$. Similarly, $\epsilon_{p}$ is the energy of the outer-shell $2p$ state, and $p_{m\sigma}^{\dag}$ is the creation operator for the $2p$ state with orbital magnetic quantum number $m$ and spin $\sigma$. $H_{S-O}$ is the spin-orbit interaction for the outer-shell $2p$ state, and $H_{exch}$ is the exchange Coulomb interaction between inner-shell $1s$ and outer-shell $2p$ electrons that contributes in the intermediate state with a $1s$-core hole. The second term in eq.(\ref{1}) is represented as
\begin{align}
\label{3}
    H_{PE}=\sum_{\sigma}\int d\varepsilon\ \varepsilon c_{\varepsilon\sigma}^{\dag}c_{\varepsilon\sigma},
\end{align}
where $c_{\varepsilon\sigma}^{\dag}$ is the creation operator for the photoelectron with the kinetic energy $\varepsilon$ and the spin $\sigma$. In this study, we assume linearly polarized light for incident X-ray. Then, the photoemission from the inner-shell $1s$ state is simple and shows a $p$-like intensity distribution along the direction of the polarization vector. Thus, we use a simple notation for $H_{PE}$ because the angular dependence is obvious and does not affect the following discussion, except for the kinetic energy and the spin state of photoelectrons.

The quantum states described by the unperturbed Hamiltonian in eq.(\ref{1}) undergoes a perturbative transition due to the electron-photon interactions,
\begin{align}
\label{4}
    V=\sum_{\sigma}\int d\varepsilon \ c_{\varepsilon\sigma}^{\dag}s_{\sigma}ae^{-i\Omega t}+M_{ph}e^{i\omega t}+h.c.
\end{align}
In eq.(\ref{4}) the first and second terms describe the emission of photoelectrons and X-ray photons, respectively, where $a$ is the annihilation operator for the incident photons with the energy $\Omega$, while $\omega$ is the emitted photon energy through the $M_{ph}$ term given in eq.(\ref{5}) below. Note that the coefficients for the terms are omitted for simplicity because they are irrelevant to the following discussions.
The operator $M_{ph}$ describes the emission of X-ray photons through the $2p\rightarrow1s$ dipole transition
and is expressed as
\begin{align}
\label{5}
M_{ph}=\sum_{\lambda,m,\sigma}\braket{s|\vec{e}_{\lambda}\cdot\vec{r}|p_{m}}b_{\lambda}^{\dag}s_{\sigma}^{\dag}p_{m\sigma},
\end{align}
where $b_{\lambda}^{\dag}$ represents the creation operator for emitted X-ray photon with the polarization $\lambda$ and $\vec{e}_{\lambda}$ is the polarization vector satisfying the Coulomb gauge condition. In the present study, we consider the linear polarization as $\lambda$.

The $2p\rightarrow1s$ XEPECS process is obtained by the coincidence measurement for energies, $\varepsilon$ and $\omega$. In the rotating-wave approximation \cite{Loudon}, the wave function of the $2p\rightarrow1s$ XEPECS process is given in the framework of the coherent second-order optical process and is expressed as
\begin{align}
\label{6}
    \ket{\psi^{(2)}(t)}=-\lim_{t_{0}\rightarrow-\infty}\int_{t_{0}}^{t}dt_{2}\int_{t_{0}}^{t_{2}}dt_{1}V_{I}(t_{2})e^{-\Gamma_{1s}(t_{2}-t_{1})}V_{I}(t_{1})\ket{\psi(t_{0})}.
\end{align}
Here, $V_{I}(t)$ in eq.(\ref{6}) means the interaction picture $e^{iH_{0}t}Ve^{-iH_{0}t}$ for $V$ in eq.(\ref{4}). An exponential decay factor with the lifetime $1/\Gamma_{1s}$ for the $1s$ core-hole states is empirically included in eq.(\ref{6}). Note that, to simplify the equation, we take the natural unit system ($\hbar=1$) and $t_{0}$ to the limit of negative infinity.
As a result, the wave function in the XEPECS process is given by 
\begin{align}
\label{7}
    \ket{\psi^{(2)}(t,\varepsilon,\omega)}=-i\sum_{f,i,\sigma,\lambda}\frac{\braket{f_{\sigma\lambda}|M_{ph}|i_{\sigma}}\braket{i_{\sigma}|c_{\varepsilon\sigma}^{\dag}s_{\sigma}a|g}}{\Omega+E_{g}-E_{i_{\sigma}}-\varepsilon+i\Gamma_{1s}}\ket{f_{\sigma\lambda}}
    \int_{-\infty}^{t}dt_{2}e^{i(E_{f_{\sigma\lambda}}+\varepsilon+\omega-E_{g}-\Omega)t_{2}}.
\end{align}
Here $\ket{g}$, $\ket{i_{\sigma}}$ and $\ket{f_{\sigma\lambda}}$ denote the initial, intermediate and final states with the energies $E_{g}$, $E_{i_{\sigma}}$ and $E_{f_{\sigma\lambda}}$, respectively.
In addition, $\sigma$ and $\lambda$ represents the spin of the photoelectron and the linear polarization of the emitted photon.
The term with time integration over $t_{2}$ vanishes in the normalization. Therefore, this term is considered to be an irrelevant factor in the present study, and thus we omit $t$ in the notation $\ket{\psi^{(2)}}$ below.

\begin{figure}[H]
    \centering
    \includegraphics[scale=0.5]{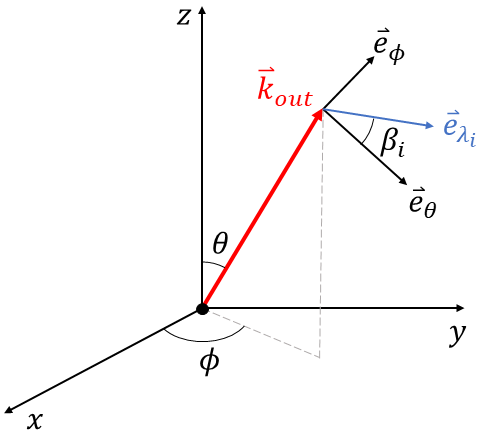}
    \caption{Geometrical setting of the emitted photon. The red line is the wave-number vector of emitted X-ray photons, while the blue one the linear polarization vector of it.}
    \label{fig:Ang of x-ray}
\end{figure}

The wave function $\ket{\psi^{(2)}(\varepsilon,\omega)}$ is calculated by diagonalizing the Hamiltonian $H_{0}$ in the vector sub-space as shown below. First, we consider the initial state. The total angular momentum $J$ of the $sp$-electron system in the initial state is $1/2$ due to the splitting of the 2$p$ state by the spin-orbit interaction. In this study, we consider the initial state with the total magnetic quantum number $M = 1/2$.
The initial state with $(J,M)=(1/2,1/2)$ is thus expressed as
\begin{align}
\label{8}
    \ket{g}=s_{\uparrow}^{\dag}s_{\downarrow}^{\dag}\biggl(\sqrt{\frac{2}{3}}p_{1\downarrow}^{\dag}-\sqrt{\frac{1}{3}}p_{0\uparrow}^{\dag}\biggl)a^{\dag}\ket{0},
\end{align}
where $a^{\dag}$ is the creation operator of an incident photon and $\ket{0}$ denotes the vacuum state. In the case of the intermediate state, $M$ of the $sp$-electron system takes the values 0 or 1 depending on the spin direction of the photoelectron, and the basis states are given by
\begin{align}
\label{9}
    \begin{cases}
    \ket{s_{\uparrow}p_{0\downarrow}},\ket{s_{\uparrow}p_{-1\uparrow}},\ket{s_{\downarrow}p_{1\downarrow}},\ket{s_{\downarrow}p_{0\uparrow}}\ &(M=0\ \rm{for} \uparrow \rm{PE})\\
    \ket{s_{\uparrow}p_{1\downarrow}},\ket{s_{\uparrow}p_{0\uparrow}},\ket{s_{\downarrow}p_{1\uparrow}}\ &(M=1\ \rm{for} \downarrow \rm{PE})
    \end{cases},
\end{align}
where PE means the photoelectron. The intermediate state has the electron configuration of $s^{1}p^{1}$ in the present situation and the total orbital- and spin-angular momentum take the values as $L = 1$ and $S = 0, 1$, which produces the coupled basis denoted by $\ket{(LS)JM}$. Then, the coupled basis provides the eigenstates of the $sp$-hamiltonian including the spin-orbit interaction and thus the eigenenergy can be labeled in terms of the total angular momentum $J = 0, 1, 2$. Actually, the basis set with $M = 0$ in eq.(\ref{9}) spans the vector space $\ket{(1S)J0}$ with $(J,S) = (0,1), (1,0), (1,1)$ and $(2,1)$, while those with $M = 1$ spans the vector space $\ket{(1S)J1}$ with $(J,S) = (1,0),(1,1)$ and $(2,1)$. The final state $\ket{f_{\sigma\lambda}}$ is given by $s_{\uparrow}^{\dag}s_{\downarrow}^{\dag}c_{\varepsilon\sigma}^{\dag}b_{\lambda}^{\dag}\ket{0}$, where $b_{\lambda}^{\dag}$ is the creation operator of the emitted photon and $(J,M)$ of the $sp$-electron system $s_{\uparrow}^{\dag}s_{\downarrow}^{\dag}\ket{0}$ is $(0,0)$. The dipole transition matrix of $M_{ph}$ between the final state and the intermediate basis state with $M$ = 0 and 1 is given by
\begin{align}
\label{10}
\left(
\begin{array}{cccc|ccc}
\braket{f_{\uparrow\lambda_{1}}|M_{ph}|s_{\uparrow}p_{0\downarrow}} & \dots & \braket{f_{\uparrow\lambda_{1}}|M_{ph}|s_{\downarrow}p_{0\uparrow}} &  &  & \text{\huge{0}} &\\
\braket{f_{\uparrow\lambda_{2}}|M_{ph}|s_{\uparrow}p_{0\downarrow}} & \dots & \braket{f_{\uparrow\lambda_{2}}|M_{ph}|s_{\downarrow}p_{0\uparrow}} &  &  &  & \\ \hline
&  &  &  &  \braket{f_{\downarrow\lambda_{1}}|M_{ph}|s_{\uparrow}p_{1\downarrow}} & \dots & \braket{f_{\downarrow\lambda_{1}}|M_{ph}|s_{\downarrow}p_{1\uparrow}}\\
& \text{\huge{0}} &  &  & \braket{f_{\downarrow\lambda_{2}}|M_{ph}|s_{\uparrow}p_{1\downarrow}} & \dots & \braket{f_{\downarrow\lambda_{2}}|M_{ph}|s_{\downarrow}p_{1\uparrow}}
\end{array} 
\right) \notag\\\equiv \left(
\begin{array}{cccc|ccc} 
   & M_{ph}^{(\uparrow\lambda)} &  &  &  & \text{\huge{0}}     &  \\ 
   &                   &  &  &  &                     &  \\ \hline
   &                   &  &  &  &                     &  \\
   & \text{\huge{0}}   &  &  &  & M_{ph}^{(\downarrow\lambda)} &
\end{array} 
\right),
\end{align}
where $M_{ph}^{(\uparrow\lambda)}$ and $M_{ph}^{(\downarrow\lambda)}$ are represented using the coefficients $\alpha^{(\lambda_{i})}_{-m}$ defined below as 
\begin{align}
\label{11}
    M_{ph}^{(\uparrow\lambda)} = 
    \left(
\begin{array}{cccc} 
  \sqrt{\frac{1}{3}}\alpha_{0}^{(\lambda_{1})} & 0 & 0 &  -\sqrt{\frac{1}{3}}\alpha_{0}^{(\lambda_{1})} \\ 
  \sqrt{\frac{1}{3}}\alpha_{0}^{(\lambda_{2})} & 0 & 0 &  -\sqrt{\frac{1}{3}}\alpha_{0}^{(\lambda_{2})} \\ 
\end{array} 
\right) \mathrm{and}\ M_{ph}^{(\downarrow\lambda)} = 
    \left(
\begin{array}{ccc} 
  -\sqrt{\frac{1}{3}}\alpha_{-1}^{(\lambda_{1})} & 0 & \sqrt{\frac{1}{3}}\alpha_{-1}^{(\lambda_{1})} \\ 
  -\sqrt{\frac{1}{3}}\alpha_{-1}^{(\lambda_{2})} & 0 &  \sqrt{\frac{1}{3}}\alpha_{-1}^{(\lambda_{2})} \\ 
\end{array} 
\right),
\end{align}
in terms of the final state $\ket{f_{\sigma\lambda}}$ and the basis set for the intermediate state described in eq.(\ref{9}).
In the Coulomb gauge conditions, there are two 
independent polarization modes $\lambda_{1}$ and $\lambda_{2}$, being perpendicular to the wave number vector $\Vec{k}_{out}$ of the emitted X-ray photon. 
In this paper, the polarization of the emitted X-ray photon is described by linear polarization, and then the elements $\alpha_{0}^{(\lambda_{i})}$ and $\alpha_{-1}^{(\lambda_{i})}$ in eq.(\rm\ref{11}) are given by
\begin{align}
\label{12}
    &\alpha_{0}^{(\lambda_{i})}=-\cos{\beta_{i}}\sin{\theta},\\
    \label{13}
    &\alpha_{-1}^{(\lambda_{i})}=\sqrt{\frac{1}{2}}(\cos{\beta_{i}}\cos{\theta}+i\sin{\beta_{i}})e^{i\phi},
\end{align}
where $\theta$ and $\phi$ are the polar coordinates for $\Vec{k}_{out}$ and $\beta_{i}$ describes the direction of polarization vector $\Vec{e}_{\lambda_{i}}$, as illustrated in Fig.\ref{fig:Ang of x-ray}. Here $\Vec{e}_{\lambda_{i}}$ is described as a liner combination of $\Vec{e}_{\theta}$ and $\Vec{e}_{\phi}$, given by 
\begin{align}
\label{14}
    \Vec{e}_{\theta}&=(\cos{\theta}\cos{\phi},\cos{\theta}\sin{\phi},-\sin{\theta}),\\
    \label{15}
    \Vec{e}_{\phi}&=(-\sin{\phi},\cos{\phi},0).
\end{align}
In this paper, we take the linear polarization of $\beta_{1}=\ang{90}$ for $\lambda_{1}$ and $\beta_{2}=\ang{180}$ for $\lambda_{2}$. As shown in eq.(\ref{11}), it is important for the formation of entanglement states that the selection rule for the X-ray emission leads to two combinations between the PE spins and the linear polarization coefficient $\alpha_{-m}$, i.e., ($\uparrow$, $\alpha_{0}$) and ($\downarrow$, $\alpha_{-1}$), where $m$ corresponds to the orbital magnetic quantum number of the 2$p$ orbital contributing to the radiative decay $2p_{m\sigma}\rightarrow1s_{\sigma}$.

Applying (\ref{8})-(\rm\ref{11}) to eq. (\ref{7}), the wave function of the whole system "just before observation" is given by
\begin{align}
\label{16}
    \ket{\psi^{(2)}(\varepsilon,\omega)}=-i\sum_{\lambda}\biggl(A_{\uparrow,\lambda}(\varepsilon)\ket{c_{\varepsilon\uparrow}b_{\lambda}}+A_{\downarrow,\lambda}(\varepsilon)\ket{c_{\varepsilon\downarrow}b_{\lambda}}\biggl)\ket{s_{\uparrow}s_{\downarrow}},
\end{align}
where $A_{\uparrow,\lambda}(\varepsilon)$ and $A_{\downarrow,\lambda}(\varepsilon)$ are 
\begin{align}
\label{17}
    &A_{\uparrow,\lambda}(\varepsilon)\equiv\sum_{i}\frac{\braket{f_{\uparrow\lambda}|M_{ph}^{(\uparrow,\lambda)}|i_{\uparrow}}\braket{i_{\uparrow}|c_{\varepsilon\uparrow}^{\dag}s_{\uparrow}a|g}}{\Omega+E_{g}-E_{i_{\uparrow}}-\varepsilon+i\Gamma_{1s}},\\
\label{18}
    &A_{\downarrow,\lambda}(\varepsilon)\equiv\sum_{i}\frac{\braket{f_{\downarrow\lambda}|M_{ph}^{(\downarrow,\lambda)}|i_{\downarrow}}\braket{i_{\downarrow}|c_{\varepsilon\downarrow}^{\dag}s_{\downarrow}a|g}}{\Omega+E_{g}-E_{i_{\downarrow}}-\varepsilon+i\Gamma_{1s}},
\end{align}
and $\ket{i_{\sigma}}$ is the intermediate state with energy $E_{i\sigma}$. 
Actually, as we will see later, these coefficients are simplified as $A_{\uparrow,\lambda}\propto\frac{1}{3}\alpha_{0}^{(\lambda)}$ and $A_{\downarrow,\lambda}\propto\frac{\sqrt{2}}{3}\alpha_{-1}^{(\lambda)}$ in the present treatment.
By normalizing the wave function in eq.(\ref{16}) and omitting the closed inner shell $\ket{s_{\uparrow}s_{\downarrow}}$, we obtain the spin-polarization entanglement state
\begin{align}
\label{19}
    \ket{\psi(\varepsilon)}_{AB}=\frac{1}{\sqrt{\sum_{\sigma,\lambda}|A_{\sigma,\lambda}|^{2}}}\sum_{\sigma,\lambda}A_{\sigma,\lambda}\ket{c_{\varepsilon\sigma}}_{A}\ket{b_{\lambda}}_{B}.
\end{align}

We consider the entanglement entropy \cite{PhysRevA.53.2046}, a function that quantifies the degree of entanglement between two subsystems A and B. In this paper, we take the spin of PE and the polarization of emitted X-ray photons as subsystems A and B, respectively. 
For the calculation of the entanglement entropy, we need to calculate first the reduced density matrix $\rho_{A}$:
\begin{align}
\label{20}
    \rho_{A}=\mathrm{Tr}_{B}[\ket{\psi(\varepsilon)}\bra{\psi(\varepsilon)}_{AB}]=\frac{1}{\sum_{\sigma,\lambda}|A_{\sigma,\lambda}|^{2}}\sum_{\sigma,\sigma',\lambda}A_{\sigma,\lambda}A_{\sigma',\lambda}^{*}\ket{c_{\varepsilon\sigma}}\bra{c_{\varepsilon\sigma'}}_{A}.
\end{align}
For the eigenvalues $x_{i}$ of $\rho_{A}$, consider the von Neumann entropy $S_{A}$ \cite{PhysRevA.53.2046}:
\begin{align}
\label{21}
    S_{A}=-\mathrm{Tr}[\rho_{A}\log_{2}\rho_{A}]=-\sum_{i}x_{i}\log_{2}x_{i},
\end{align}
which we regard as the entanglement entropy in the present study.
This quantity provides the degree of entanglement between the spin of PE and the linear polarization of the emitted X-ray photon.

\section{Results and discussion}

We show the numerical results of the density matrix and entanglement entropy of the $sp$-model. The values of the parameters used in the calculations are listed in Table \ref{table:data_type}. Here $G$ is strength of the exchange Coulomb interaction, $\zeta$ is the spin-orbit coupling constant, $\epsilon_{s}$ and $\epsilon_{p}$ are $1s$ and $2p$ energy levels, $\Omega$ is the incident photon energy, $\Gamma_{1s}$ is the lifetime of the $1s$ inner-shell hole in the intermediate state, and $\gamma$ is the resolution for the emitted photon energy.

\begin{figure}[H]
    \centering
    \includegraphics[scale=0.5]{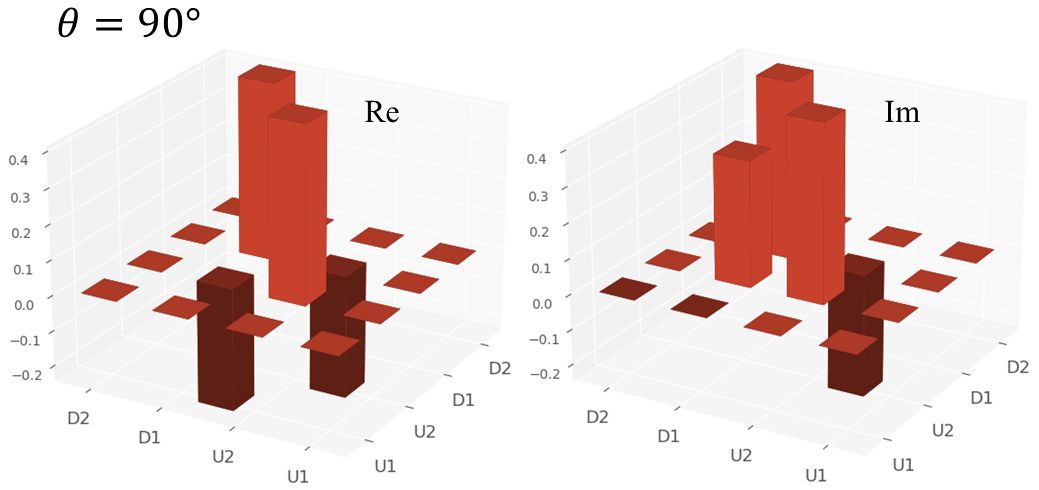}
    \caption{Real (Re) and imaginary (Im) parts of the density matrix for spin-polarization entanglement states. The basis used here is the combined states between photoelectron spin (U for up and D for down) and emitted photon polarization (1 for $\lambda_{1}$ and 2 for $\lambda_{2}$).}
    \label{fig:density matrix}
\end{figure}

\begin{table}[H]
    \centering
    \scalebox{1.0}[1.05]{
    \begin{tabular}{ccccccc}
         \hline \hline
        $G$ & $\zeta$ &  $\epsilon_{s}$ & $\epsilon_{p}$ & $\Omega$ & $\Gamma_{1s}$ & $\gamma$ \\
        \hline
        0.3  &  0.1  &  -13.6 & -5.0 & 20.0 & 0.5 & 0.4 \\
        \hline \hline
    \end{tabular}
    }
    \caption{Parameter values used in the $sp$-model calculation in the unit of eV.}
    \label{table:data_type}
\end{table}
First, we consider the results of the density matrix of the $sp$-model. In Fig.\ref{fig:density matrix}, we show the calculated real and imaginary parts of the density matrix at the emission angle $\theta=\ang{90}$. Here U and D denote the photoelectron up and down spin state, and 1 and 2 the linear polarization $\lambda_{1}$ and $\lambda_{2}$ of the emitted X-ray photon. We used the coupled states between spin and polarization as the bases for the calculation of the density matrix. 
In the case of the $sp$-model, the optical selection rule in eq.(\ref{11}) causes the disappearance of matrix components of $\ket{U1}$ and $\ket{D2}$ as discussed above, and leaves only the components of $\ket{U2}$ and $\ket{D1}$.
The off-diagonal terms $\ket{U2}\bra{D1}$ and $\ket{D1}\bra{U2}$ provide the coherence of the spin and polarization between the photoelectron and the emitted X-ray photon.

Fig.\ref{fig:density matrix} shows that eq.(\ref{19}) at $\theta=\ang{90}$ is the maximum entanglement state. In fact, the entanglement entropy in eq.(\ref{21}) takes the value 1.
According to eq.(\ref{19}), the spin-polarization entanglement state is given by
\begin{align}
\label{22}
    \ket{\psi(\varepsilon)}_{AB}=\frac{1}{\sqrt{C}}(A_{\uparrow,\lambda_{2}}\ket{c_{\varepsilon\uparrow}}_{A}\ket{b_{\lambda_{2}}}_{B}+
    A_{\downarrow,\lambda_{1}}\ket{c_{\varepsilon\downarrow}}_{A}\ket{b_{\lambda_{1}}}_{B}),
\end{align}
where $C$ is defined as  $\sum_{\sigma,\lambda}|A_{\sigma,\lambda}|^{2}$ for the normalization.
In this case the coefficients $A_{\uparrow,\lambda_{2}}$ and $A_{\downarrow,\lambda_{1}}$ become equal at $\theta=\ang{90}$, leading a maximum entanglement state.

\begin{figure}[H]
    \centering
    \includegraphics[scale=0.5]{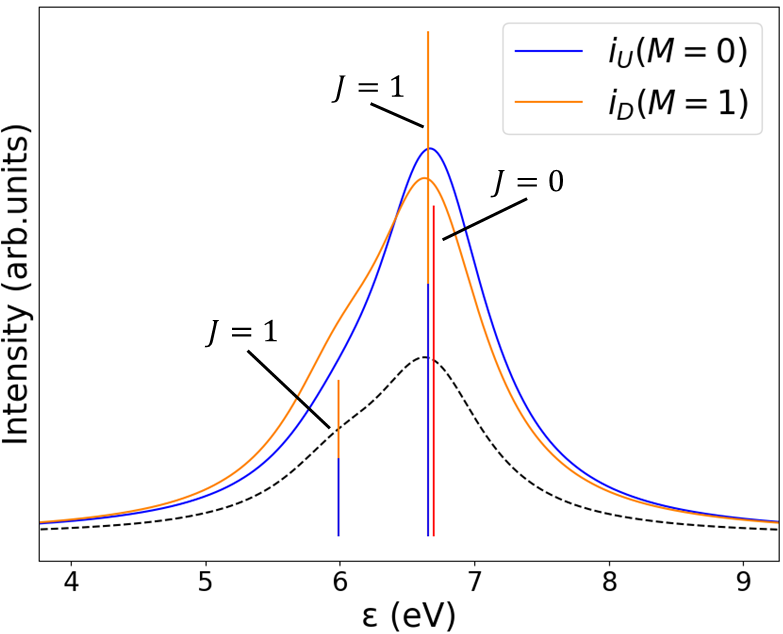}
    \caption{The calculated spin-resolved 1s XPS. The horizontal axis represents the kinetic energy of the photoelectron 
and the vertical axis is the photoelectron intensity. The up and down spin spectra are shown by the blue $i_{U}$ with $M = 0$ and orange $i_{D}$ with $M = 1$ lines, respectively. The blue and red vertical bars are the line spectra for the up-spin, while the orange vertical bars are the line spectra for the down-spin. The black dotted line is the spectrum of up-spin except for the red vertical bar.}
    \label{fig:1sXPS}
\end{figure}

The discussion above can be visualized in the 2$p\rightarrow$1$s$ XEPECS calculation. Before discussing the XEPECS results, we first consider the spin-resolved 1$s$ XPS shown in Fig.\ref{fig:1sXPS}.
In this figure, the 1$s$ XPS intensities for up- and down spin are shown by the blue and orange curves, respectively, as a function of kinetic energy. Here, the continuous spectra are obtained by the convolution of the vertical bars calculated from the XPS function, 
\begin{align}
\label{23}
    F_{\rm XPS}(\varepsilon,\sigma)=\sum_{i}|\braket{i_{\sigma}|c_{\varepsilon\sigma}^{\dag}s_{\sigma}a|g}|^{2}\delta(\Omega+E_{g}-E_{i_{\sigma}}-\varepsilon) ,
\end{align}
with the Lorentz function of the width of $\Gamma_{1s} = 0.5$ eV.
We observe the clear spin-dependence in 1$s$ XPS and the difference is simply understood as follows.
Since the initial state has the character $(J,M) = (1/2, 1/2)$, only the $J = 0$ and $1$ XPS final states are allowed through the dipole emission of the 1$s$ electron, hence the transition to the $J = 2$ states included in the present basis set in eq.(\ref{9}) becomes forbidden. The $M$ values of the XPS final states are $0$ and $1$ for up and down spin emission, respectively. Thus, the two $J = 1$ states, which are given by the linear combination between $\ket{(10)1M}$ and $\ket{(11)1M}$ states, appear in both up- and down-spin XPS, and these two lines degenerate between $M = 0$ and $1$ about the kinetic energy. The $J = 0$ state, however, appears only in the up-spin XPS with $M = 0$.

\begin{figure}[H]
    \centering
    \includegraphics[scale=0.5]{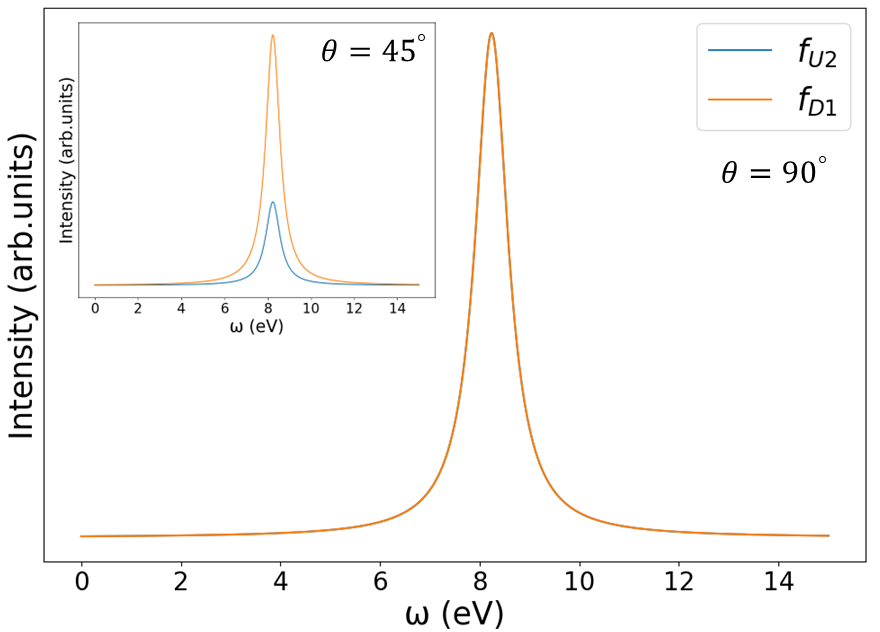}
    \caption{The results of 2$p\rightarrow$1$s$ XEPECS calculated at $\theta=\ang{90}$ and $\ang{45}$ (inset). The horizontal axis means the emission energy and the vertical axis is the emission intensity. In the result of $\theta=\ang{90}$, the spectra corresponding to U2 and D1 completely overlap with each other. On the other hand, the spectra at the $\theta=\ang{45}$ show a clear difference between U2 and D1.}
    \label{fig:2p-1sXEPECS}
\end{figure}

Next, we discuss the calculated 2$p\rightarrow$1$s$ XEPECS shown in Fig.\ref{fig:2p-1sXEPECS} where the intensities of the emitted photon are plotted for the two-cases $f_{U2}$ and $f_{D1}$ against the emitted photon energy. In this calculation, the photoelectron kinetic energy is set at 6.66 eV of the spin-resolved $1s$ XPS shown in Fig.\ref{fig:1sXPS}. In the case of emission angle $\theta=\ang{45}$ of the inset, the emission intensities of $f_{U2}$ and $f_{D1}$ are different from each other, but in the case of $\theta=\ang{90}$, the intensities of the two spectra are the same, which means that the degree of entanglement depends on the emission angle $\theta$ and becomes maximum at $\theta=\ang{90}$.

\begin{figure}[H]
    \centering
    \includegraphics[scale=0.6]{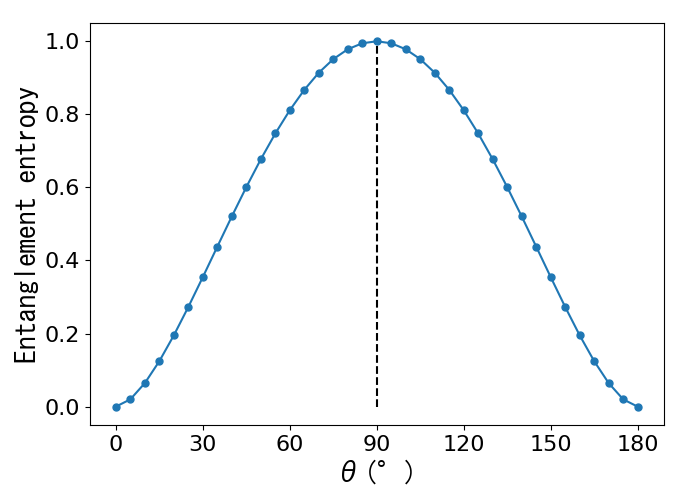}
    \caption{The photon emission-angle dependence of the entanglement entropy. The black dashed line is for eye guide for $\theta=\ang{90}$.}
    \label{fig:Ang of entangle}
\end{figure}

Finally, we discuss the emission angle dependence of the entanglement entropy $S_{A}$. Fig.\ref{fig:Ang of entangle} shows the $\theta$ dependence of the entanglement entropy $S_{A}$ in the range of $\ang{0}\leq\theta\leq\ang{180}$. 
The figure shows that $S_{A}$ has a maximum value of 1 at $\theta=\ang{90}$ and monotonically decreases as $\theta$ varies from $\theta=\ang{0}$ to $\theta=\ang{180}$. This behavior of $S_{A}$ can be explained directly from eq.(\ref{19}). We first note that only the intermediate states with $J = 1$ contribute to the coefficient $A_{\sigma,\lambda}(\varepsilon)$ in eq.(\ref{17}) and (\ref{18}). As mentioned above, there are eigenstates with $J = 0$, $1$, and $2$ in the $s^{1}p^{1}$ configuration for the intermediate states. Among these, the initial state is excited into $J = 0$ and $1$ states as shown in the spin-resolved 1$s$ XPS in Fig.\ref{fig:1sXPS}. On the other hand, the radiative decay from the $J = 0$ state to the final state with $J = 0$ is forbidden because of the dipole selection rule, or because of the forbidden spin-state change from the intermediate state $S = 1$ to the final state $S = 0$. Thus, only the intermediate state with $J = 1$ can contribute to the coefficients $A_{\sigma,\lambda}(\varepsilon)$, and they are reduced to $A_{\uparrow,\lambda} \propto \frac{1}{3}\alpha_{0}^{(\lambda)}$ and $A_{\downarrow,\lambda} \propto \frac{\sqrt{2}}{3}\alpha_{-1}^{(\lambda)}$ as discussed below.

In the present study, we have two eigenstates with $J = 1$, which are linear combinations between $(J,S) = (1,0)$ and $(1,1)$ states for both $M = 0$ and $M = 1$ intermediate state, and the corresponding lines in 1$s$ XPS degenerate between $M = 0$ and $M = 1$.
Furthermore, the transition strength to the $J = 1$ states of $M = 1$ is just two times larger than that in the $M = 0$ case, as can be seen from the comparison of $i_{D}$ curve with the dashed curve which is obtained by convoluting only $J = 1$ vertical lines. In this way, the denominators in eq.(\ref{17}) and (\ref{18}) become the same for the terms under the summation, and the numerators are given by geometric factors provided by the Wigner – Eckart theorem, which leads to simple expressions of $A_{\uparrow,\lambda} = \frac{1}{3}\alpha_{0}^{(\lambda)}$ and $A_{\downarrow,\lambda} = \frac{\sqrt{2}}{3}\alpha_{-1}^{(\lambda)}$, except a common factor. 
Using this fact and $\beta_{1},\beta_{2} = \ang{90}, \ang{180}$, the eq.(\ref{19}) is simplified and the spin-polarization entanglement state is represented as 
\begin{align}
\label{24}
\ket{\psi(\varepsilon)}_{AB}=\frac{1}{3\sqrt{C}}\biggl[\sin{\theta}\ket{c_{\varepsilon\uparrow}}_{A}\ket{b_{\lambda_{2}}}_{B}
+e^{i\phi}\bigg(i\ket{c_{\varepsilon\downarrow}}_{A}\ket{b_{\lambda_{1}}}_{B}
-\cos{\theta}\ket{c_{\varepsilon\downarrow}}_{A}\ket{b_{\lambda_{2}}}_{B}\bigg)\biggl],
\end{align}
which is a general form of the $sp$-model including the case with $\theta \neq \ang{90}$. 
Furthermore, this expression explains the fact that the entanglement entropy becomes constant against the kinetic energy of the photoelectrons for any emission angle in the present calculation, i.e., the coefficients for the basis kets are independent of the kinetic energy of the photoelectrons.

From Fig.\ref{fig:Ang of entangle}, we see that $S_{A}$ decreases as $\theta$ increases or decreases from $\ang{90}$. This is due to the increase of the value of the coefficient $-\frac{1}{3\sqrt{C}}e^{i\phi}\cos{\theta}$ in the third term of eq.(\ref{24}). In addition, the spin-polarization entanglement disappears at $\theta = \ang{0}$ and $\ang{180}$. This is because the first term coefficient $\frac{1}{3\sqrt{C}}\sin{\theta}$ becomes zero, and thus eq.(\ref{24}) includes only the down-spin terms $\ket{c_{\varepsilon\downarrow}}_{A}\ket{b_{\lambda_{1}}}_{B}$ and $\ket{c_{\varepsilon\downarrow}}_{A}\ket{b_{\lambda_{2}}}_{B}$. Here, we emphasize that the feature obtained in Fig.\ref{fig:Ang of entangle} does not depend on how to set the polarization $\lambda_{1}$ and $\lambda_{2}$, i.e., we can obtain the same result for any choice of orthogonal linear polarization, owing to the invariance of the trace in eq.(\ref{20}) for the orthogonal transformation of the basis in the subspace B.

The mechanism for the emergence of spin-polarization quantum entanglement is due to the sequence of quantum coherent optical processes from $1s$ inner-shell excitation to radiative relaxation of $2p$ state.
This sequence of events can be divided into two steps. In the first step, the $1s$ inner-shell electrons are excited as photoelectrons by the incident photon. In the second step, during the radiative relaxation of the $2p$ electron into the $1s$ inner-shell hole, the $2p$ spin-orbital component projected onto the spin of the $1s$ inner-shell hole is "transferred" to the polarization component of the emitted X-ray photon.
The point is the spin-orbit interaction acting on the $2p$ electrons. This interaction affects the polarization component of the emitted X-ray photon depending on the spin of the $1s$ inner-shell hole. In this way, the entanglement occurs between the spin of the photoelectron and the linear polarization of the emitted X-ray photon.

\section{Conclusion}
We have shown using the $sp$-model that the spin-polarization entanglement occurs between photoelectrons and emitted X-ray photons in 2$p\rightarrow$1$s$ XEPECS process and found that the degree of entanglement depends on the emission angle of the photons. 

We showed that the degree of entanglement of the up spin/$\lambda_{2}$ and down spin/$\lambda_{1}$ pairs, which occurs from the spin-orbit interaction acting on the 2$p$ orbitals, takes a maximum value of 1 at the emission angle $\ang{90}$ and a minimum value of 0 at $\ang{0}$ and $\ang{180}$. This indicates that the manner of entanglement depends on at which angle the emitted X-ray photons are detected. 
Actually, such angular dependence of the degree of spin-polarization entanglement have been pointed out by B.N.Chowdhury $et\ al.$\cite{mainref} in high-energy Compton scattering. It should be noted that the Compton scattering is a first-order optical process, while the XEPECS process is a second-order one and thus the mechanism of the emergence of the entanglement and its angular dependence is different in both cases.
Thus, our results provides a new knowledge about the quantum entanglement in high-energy region. We consider that increasing of such knowledge related to quantum entanglement in the X-ray region will enable us to open new fields.

Finally, we would like to emphasize that the present theory is positioned as research in the fields that explicitly treat the quantized state of light, such as quantum optics and quantum information. The use of a quantum theory of light, beyond conventional semiclassical treatments, is expected to lead to a discovery of new physical phenomena related to the interaction between the electronic state of solids and light.

\begin{acknowledgment}
%\acknowledgment

We would like to thank Professor Goro Oohata and Professor Satoshi Tanaka for valuable discussions. This work was supported by JST, the establishment of university fellowships towards the creation of science technology innovation, Grant Number JPMJFS2138.

\end{acknowledgment}

\end{document}